\documentclass[journal]{IEEEtran}
\ifCLASSINFOpdf
\else
\fi

\usepackage[colorlinks, citecolor=blue]{hyperref}
\usepackage{booktabs}
\usepackage{subcaption}
\usepackage{setspace, amsmath, amssymb, url, lscape,  algorithmic, multirow, pslatex, listings, verbatim, alltt, amsfonts, wrapfig, boxedminipage, color, cite, bookmark, pifont} %algpseudocode
\usepackage{caption} 
\usepackage[vlined,linesnumbered,ruled,boxed]{algorithm2e}
\usepackage[dvips]{graphicx}
\usepackage{epsfig}

\usepackage{xcolor}
\usepackage{balance}
\newcommand{\qed}{\nobreak \ifvmode \relax \else
      \ifdim\lastskip<1.5em \hskip-\lastskip
     \hskip1.5em plus0em minus0.5em \fi \nobreak
      \vrule height0.75em width0.5em depth0.25em\fi}

\newcommand{\paperfolder}{.}

\newcommand{\eg}{{\it e.g., }}
\newcommand{\etal}{{\it et~al. }}
\newcommand{\ie}{{\it i.e., }}

\newcommand{\comments}[1]{}
\newcommand\hl{\bgroup\markoverwith
  {\textcolor{yellow}{\rule[-.5ex]{2pt}{2.5ex}}}\ULon}

\newlength{\boxfigwidth}

\newcommand{\boxfig}[1]{
\begin{figure}[h]
\begin{center}
\begin{small}
\setlength{\boxfigwidth}{3.15in}
\addtolength{\boxfigwidth}{0in}
\noindent\framebox{\quad\begin{minipage}{\boxfigwidth}
#1
\vspace{-15pt}
\end{minipage}\quad}
\end{small}
\end{center}
\end{figure}
}

\begin{document}
%
%\title{Non Threshold-Based Task Dropping Strategy to Improve Robustness for Heterogeneous Computing System}
%\title{Robustness Improvement Techniques on Machine Queue after Task Mapping on Heterogeneous Computing System}
\title{Autonomous Task Dropping Mechanism to Achieve Robustness in Heterogeneous Computing Systems}
% \title{How to Make Optimal Task Dropping Decisions in a Heterogeneous Computing System?}
% \title{Robustness against Task Execution Time and Arrival Uncertainties in a Heterogeneous Computing System}
\author{
Ali Mokhtari, Chavit Denninnart, Mohsen Amini Salehi 
\\ High Performance Cloud Computing (HPCC) Laboratory,\\ School of Computing and Informatics, University of Louisiana at Lafayette, USA 
\\  \{ali.mokhtari1, chavit.denninnart1, mohsen.aminisalehi\}@louisiana.edu 

}
\maketitle              % typeset the header of the contribution
\begin{abstract}
Robustness of a distributed computing system is defined as the ability to maintain its performance in the presence of uncertain parameters. 
Uncertainty is a key problem in heterogeneous (and even homogeneous) distributed computing systems that perturbs system robustness. Notably, the performance of these systems is perturbed by uncertainty in both task execution time and arrival.
Accordingly, our goal is to make the system robust against these uncertainties. Considering task execution time as a random variable, we use probabilistic analysis to develop an autonomous proactive task dropping mechanism to attain our robustness goal. Specifically, we provide a mathematical model that identifies the optimality of a task dropping decision, so that the system robustness is maximized. Then, we leverage the mathematical model to develop a task dropping heuristic that achieves the system robustness within a feasible time complexity. Although the proposed model is generic and can be applied to any distributed system, we concentrate on heterogeneous computing (HC) systems that have a higher degree of exposure to uncertainty than homogeneous systems. Experimental results demonstrate that the autonomous proactive dropping mechanism can improve the system robustness by up to 20\%.% while reducing the incurred cost of using resources. 

\end{abstract}

\begin{IEEEkeywords}
Heterogeneous Computing (HC) Systems, Uncertainty, Dropping Mechanism, Robustness, Mapping Heuristic \end{IEEEkeywords}

\section{Introduction}
\label{sec:intro}
\subsection{Problem Statement}
Heterogeneous Computing (HC) systems can be categorized as consistent or inconsistent~\cite{li2018cost,ipdps19} heterogeneous systems. Consistent machine heterogeneity describes a computing system of multiple machines with the same architecture but different performance characteristics. In an inconsistent HC system, machines are also distinguished by their different architectures  \cite{zhao2017fpga,hong2017gpu,Grasso14}. In such a system, each task may have different execution times on different machines of the system. Formally, an \emph{inconsistently heterogeneous} system is defined as a computing system in which machine A may be faster than machine B for task 1 but slower than other machines for task 2 \cite{salehi2016stochastic}. As a popular example of an inconsistent HC system, we can consider Amazon cloud \cite{aws} that offers various machine types (\eg CPU-Optimized, Memory-Optimized, and GPU).

In the same way, task requests can be categorized as consistently or inconsistently heterogeneous. 
For instance, a system dedicated for video transcoding \cite{TPDS17} receives categorically different tasks (\ie task types) to change video resolution, bit rate, or compression formats \cite{li2018cost}. Each instance of these task types can process a video with a different size, which represents consistent heterogeneity across tasks of the same type. Such variety of tasks are proven to benefit from utilizing an HC system \cite{li2018cost}. %Typically, an HC system contains both consistent and inconsistent heterogeneity on both machine types and task types \cite{TPDS17,smith09}.%cite removed {}

\emph{Robustness} of a system is defined as its ability to maintain its performance in the face of uncertainty \cite{salehi2016stochastic,hansen2014heuristics}. Two major uncertain parameters that affect robustness of a computing system in an inconsistent HC system are, namely task execution time and task arrival \cite{TPDS17}. There is uncertainty in execution times of different task types across different machine types. Uncertainty in tasks' arrival can lead to \emph{oversubscription} situation, which is defined as an overloaded system that cannot complete all tasks by their deadlines \cite{salehi2016stochastic}.  

Co-occurrence of both tasks' arrival and execution time uncertainties in a system with inconsistent heterogeneity in their tasks and machines leads to poor resource allocation decisions and lack of robustness \cite{oxley2013energy,young13}. This is particularly crucial when resources are not abundant (\eg in Edge computing \cite{razinhpcc19}) or the resources cannot be acquired due to budget constraints (\eg in Cloud environment) \cite{TPDS17,bi2017application}. 
Accordingly, the problem we investigate in this research is: \emph{how to make an inconsistent HC system robust against uncertainties in tasks' execution times and arrival?}

\subsection{Solution Statement and Contributions}
\begin{figure} 
  \centering
  \includegraphics[width=0.4\textwidth]{\paperfolder/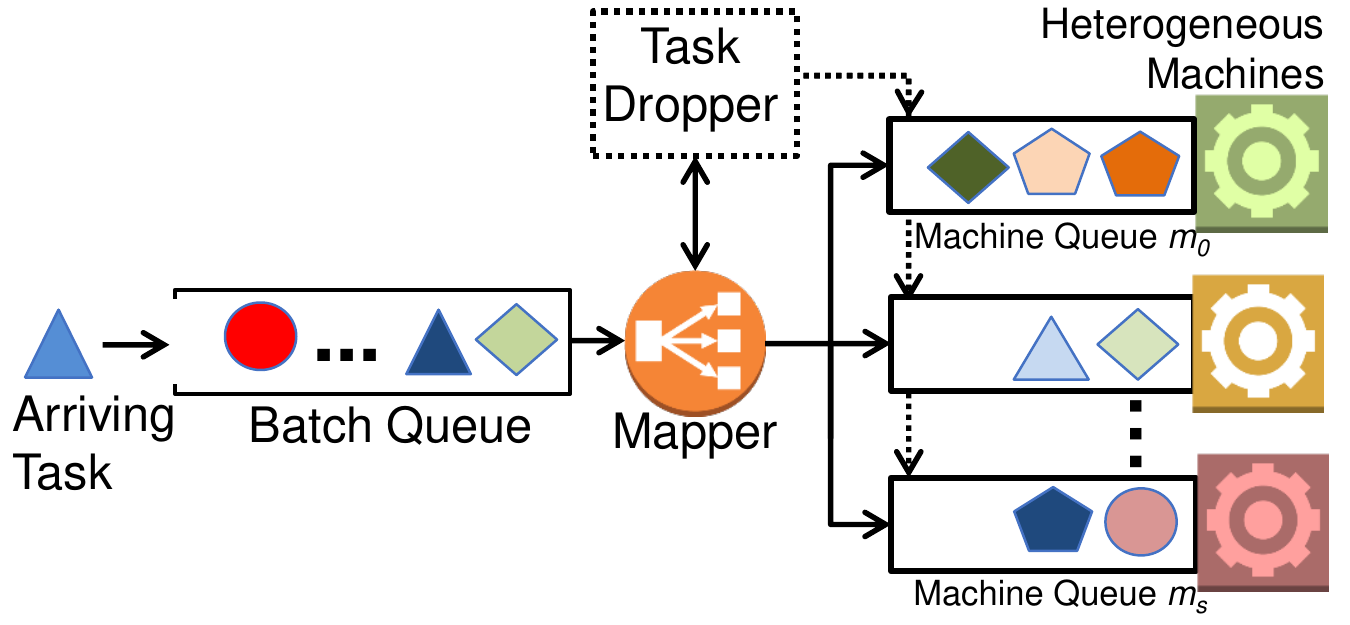}
  \caption{Overview of a batch-mode resource allocation system in a heterogeneous computing system. Task Dropper mechanism, in cooperation with the Mapper module, proactively drops tasks from machine queues to maximize the system robustness. \label{fig:overview} }
  \vspace{-10px}
\end{figure}

We address the research question in the context of an HC system used for live video streaming (\eg \cite{TPDS17,li2018cost,matin_paper}). As shown in Figure~\ref{fig:overview}, we consider an online (dynamic) batch scheduling system \cite{oxley2014makespan} to allocate tasks to heterogeneous machines. Each machine has a limited local queue (termed machine queue) to fetch data for allocated tasks before starting execution. We consider each task in the system as independent and with an individual hard deadline. Then, we measure robustness of the system based on the number of tasks completed on-time within a given time period. 

To capture the uncertainty in tasks' execution times, we model the execution times using statistical distributions and leverage them to calculate the likelihood of on-time completion for each task. Also, to capture the uncertainty in tasks' arrival rate, we utilize a task dropping mechanism that proactively drops (\ie discards) tasks that are unlikely to complete on time. Smart dropping of unlikely-to-succeed tasks not only reduces the incurred cost of using resources, but also increases the chance of success for the remaining tasks and improves the overall system robustness. However, the challenge is in making appropriate task dropping decisions to achieve the robustness goal. To address this challenge, in this work, we propose a mathematical model that at any mapping event determines the optimal task dropping decision, so that the overall system robustness is maximized. Next, we leverage the mathematical model to develop a proactive task dropping heuristic with a feasible time complexity that works along with the mapping heuristic (see Figure~\ref{fig:overview}). Although we target HC systems, the proposed model is generic and can improve the robustness of homogeneous systems too.

Prior probabilistic task dropping approaches (\eg \cite{khemka2015utility, ipdps19, denninnart2019improving}) base their dropping decisions on the chance of completing a task before its deadline (termed \emph{chance of success}) and comparing that against a user-defined threshold. Nonetheless, dropping threshold is a dynamic parameter depending on system level factors, such as task arrival intensity \cite{ipdps19}. Such a fine-grained parameter cannot be predetermined and statically applied to the HC system. Alternatively, our proposed dropping mechanism does not rely on any predefined threshold. It can autonomously make optimal dropping decisions such that the overall system robustness is maximized. 
% Experimental results indicate that our proposed task dropping mechanism outperforms prior works, both in terms of the achieved robustness and the incurred cost of using resources.

In summary, the contributions of this paper are as follows:
\begin{itemize}
% \item Evaluating properties and relation of tasks robustness in machine queue with probabilistic task dropping.
\item Developing a mathematical model for optimal proactive task dropping in an HC system.
\item Proposing an autonomous proactive task dropping heuristic in HC systems.
\item Analyzing the impact of task dropping mechanism on the robustness of both heterogeneous and homogeneous systems under varying workload characteristics.
\item Analyzing the cost benefit of using the proactive task dropping heuristic.
\end{itemize}

The rest of the paper is organized as follows: 
Section \ref{sec:background} surveys prior research works related to this research. In Section \ref{sec:sysmodel}, we present an overview of the system and our approach. Then, in Section \ref{sec:CompletionTime} we describe our mathematical model and proactive task dropping heuristic. Next, in Section \ref{sec:evltn}, performance evaluation is elaborated. We conclude the paper and provide potential future works in Section \ref{sec:conclsn}.

%\subsection{goals of approximate computing}
% energy consumption
%cost
%QoS

%=oversubscription
%In an ideal cloud computing system, all the resources are scalable without limit. Such elasticity enables all tasks to be mapped to their best affinity~\cite{li2018cost} resources at all times. However, in the real world, such an abundant resource pool is rarely achievable. Heterogeneous computing platforms frequently have limited resources to utilize. The shortage of resources can either be a limitation of the system (\eg system cannot scale up) or from the economic concerns (\ie unable to afford more resources). On moderate load, some certain resources can be busier than the other. Therefore some tasks are mapped to the resources that have less affinity but more availability. 

\section{Related Works}\label{sec:background}
In spite of substantial exploration of uncertainty in different areas, ranging from biology to economics, it has not yet been sufficiently explored in the distributed computing literature. Majority of current studies in scheduling assume a static deterministic execution environment \cite{stratus18,Marahatta19} %cite removed{Oikonomou18}
or consider predictable and stable performance for distributed computing environments \cite{Chen17,rw_11,Marahatta19}.  %cite removed{rw_12}
In practice, these assumptions do not hold. Even in the case of clouds that guarantee a certain characteristics (\eg processor speed and memory capacity) for their services, the actual performance is subject to several underlying factors, such as multi-tenancy, that cause uncertainty. To offer robustness, uncertainty and dynamic performance variations, inherent to heterogeneous and shared infrastructures \cite{Tchernykh15}, must be captured. 

Optimal task mapping in HC systems and in the presence of uncertain (stochastic) parameters has shown to be an NP-complete problem \cite{Ibarra77}. %cite removed{coffman76}
Therefore, a large body of research works has been undertaken to capture the stochastic behavior and provide a near-optimal task mapping to fulfill various performance goals (\eg minimizing average waiting time \cite{kumbhare2015reactive} 
and maximizing throughput \cite{oxley2014makespan,alba13}). %In this section, we review some of these research works that are related to this study and position our work with respect to them.

With respect to capturing uncertainty in tasks' execution time, Aupy \etal \cite{aupy2019reservation} treat tasks' execution time as a random variable and use probabilistic distributions to model the uncertainty. With the goal of minimizing the incurred cost of using cloud-based reservation, they leverage their proposed strategy to allocate an optimal reservation sequence and schedule tasks on the reserved resources. %ALI: you want to add some stuff here??

Shestak \etal~\cite{Shestak08} investigate and prepare a foundation work for stochastic task execution time modeling using probability mass function (PMF). They establish fundamental tools for the system that use PMF instead of scalar values for task scheduling. Our work builds on top of their findings, adopt their PMF modeling, calculate tasks' completion time based on convolution of PMFs, and measure robustness in a similar way to their work. 

Khemka \etal~\cite{khemka2015utility} design and evaluate four resource allocation heuristics in oversubscribed HC systems. These heuristics include the use of different utility functions based on urgency, priority, and utility class. Although they utilize PMF-based task execution times, they treat tasks' execution time in a deterministic (\ie not probabilistic) manner. Their approaches include the use of preemptive task dropping procedure (\ie discard task before reaching its deadline). However, their approach relied on a static threshold and only drop tasks, if the task's utility goes below the specified threshold.

Salehi \etal \cite{salehi2016stochastic} mathematically model the impact of task dropping on completion time PMF of tasks in an HC system. However, task dropping is carried out either based on a static threshold or in a reactive manner (\ie after a task misses its deadline). Later, Gentry \etal~\cite{ipdps19} extend the earlier study and presented a task pruning mechanism for HC systems. %With the hope that the mapping can be more favorable in the next mapping event, task deferring postpones mapping of tasks with low chance of success. 
Denninnart \etal~\cite{denninnart2019improving}, show a generalized form of the pruning mechanism and deployed it as a separate component in the system to improve robustness of homogeneous or heterogeneous systems. The generalized pruning mechanism can work in conjunction with any mapping heuristic to improve the system robustness. Nonetheless, in all of these works probabilistic task pruning make their decisions based on a predefined threshold, which is not necessarily optimal and requires user intervention. Alternatively, the dropping mechanism of this study is optimal and autonomous, \ie it does not require any predefined threshold and/or user intervention.
\section{System Model}
\label{sec:sysmodel}
This research is motivated by an inconsistent HC system used for transcoding live video streaming tasks, such as those explored in \cite{TPDS17,li2016vlsc,matin_paper}. In this system, each task has an individual deadline and it has to complete before the deadline. There is no value in executing tasks that have missed their deadlines and such tasks should be dropped to maintain liveness of the video streaming. In this HC system, a limited number of task types (\eg transcoding types) are processed.
Figure~\ref{fig:overview} shows that arriving tasks are batched in a queue; then each task is mapped to one of the $s$ heterogeneous machines.

There is uncertainty in execution time of each task type across different machine types. Furthermore, there is uncertainty in execution time of even one task type on a single machine type, due to factors such as tasks' data sizes and/or resource contention in a multi-tenant system \cite{li2016high}.
We consider the uncertain execution time of each task type as a discrete random variable and use a Probability Mass Function (PMF) to model it. Practically, execution time PMF of task type $i$ on machine type $j$ can be learned and estimated from the historic execution time information of that task type on that machine type. In an HC system, a matrix, called Probabilistic Execution Time (PET) \cite{salehi2016stochastic}, is employed to store the execution time PMFs of all task types on all machine types. Since there are limited number of known task types and machine types, the PET matrix has a limited size. It is assumed that the PET matrix is available in the HC system.

A mapping event is triggered by completing or arrival of a task to assign unmapped tasks from the batch queue. At each mapping event, first, pending tasks in machine queues that missed their individual deadlines are dropped. Then, Mapper uses a \emph{mapping heuristic} to assign unmapped tasks to available slots in machine queues. The mapping heuristic creates a temporary queue of machine-task mappings and the \emph{completion time} PMF of each unmapped task on heterogeneous machines is calculated (see Section \ref{subsec:compl}). Machine queues are to fetch data (\eg video content) for the assigned tasks, prior to their execution. To restrain the combined effect of execution times uncertainties on a task completion time uncertainty, the size of machine queues are considered to be limited.

We assume that the mapped tasks cannot be remapped, due to the data transfer overhead, and machine queues operate in a first come first serve manner. Similar to \cite{TPDS17}, tasks are considered to be sequential, independent, and executed in isolation, with no preemption and no multitasking. 

Although our model is generic and can be applied to homogeneous systems, in this study, we concentrate on HC systems. The reason is that HC systems have a higher degree of exposure to uncertainty than homogeneous systems. In fact, an inconsistent HC system is not only exposed to uncertainty in execution time of a certain task type on a given machine type, but it is also exposed to uncertainty of the same task type across different machine types. 

%Explain variation of scheduling systems, namely: those for homogeneous, immediate-mode, and batch-mode scheduling. Say because of proven higher performance, we base our analysis on batch-mode but evaluate others too. its respective scheduling 

%\input{Sources/dropping}
\section{Proactive Task Dropping}\label{sec:CompletionTime}
\subsection{Overview}
Probabilistic task dropping is a double-edged sword for system robustness. On the one hand, we miss the chance of completing a task, hence, it reduces the robustness. On the other hand, dropping improves the chance of success for the tasks behind the dropped task, as they can begin their execution earlier. To attain the maximum robustness, these two effects should be considered for any dropping decision. 
In this section, we resolve this issue and determine how task dropping decision should be made in an HC system so that the robustness is maximized. 

In essence, a task should be dropped, if it increases the likelihood of having more tasks completed on time. Therefore, in this section, firstly, we introduce a method to calculate the impact of a task dropping on the chance of success for the remaining tasks. Then, we provide a mathematical model that, at each mapping event, determines the optimal subset of tasks whose dropping can potentially maximize the robustness. %As the time complexity of provided model is not polynomial, TODO: LATER
Lastly, as the provided model is complex, we leverage it to present a sub-optimal task dropping heuristic that makes dropping decision for each individual task, as opposed to collectively considering all tasks.

\subsection{Calculating Chance of Success in Reactive Task Dropping}\label{subsec:compl} 
To calculate the chance of success for a task of type $i$ on the local queue of a machine of type $j$, we first need to determine the stochastic completion time of the task.
Recall that, the \emph{execution time} of task type $i$ on machine type $j$ is considered as a discrete random variable, denoted $E_{ij}$, which is maintained in form of a PMF in the PET matrix. Let $e_{ij}(t)$ an impulse in the PMF, representing the probability that task type $i$ on machine type $j$ takes $t$ time units to execute (\ie $e_{ij}(t)=\mathbb{P}(E_{ij}=t)$).
Similarly, let $C_{ij}$ be a discrete random variable, representing the \emph{completion time} of task type $i$ on machine $j$ and its PMF is denoted as $c_{ij}(t)$.

% Explain how to calculate completion time
%The completion time of a task arriving at time $t_s$ on an idle machine $j$  is  calculated by shifting its execution time distribution by the amount of $t_s$. 
%Then, its completion time PMF at current time $t_c$ greater than start time is determined based on Equation~\ref{eq:0}:
%\begin{equation} \label{eq:0}
%    c^{exec}_{ij}(t) = \frac{e_{ij}(t-t_s)}{ \sum_{t-t_s>t_c}^{\infty} %e_{ij}(t-t_s) } 
%\end{equation}
%\vspace{2pt}
%Here, $e_{ij}(t-t_s)$ is the shifted execution time distribution by start time. Later, it is normalized by the summation of all impulses at times greater than current time. 
% We can add a figure here to illustrate the completion time calculation of currently executing task

%The completion time of a pending task $i$ on the given machine $j$, $C^{pend}_{ij}$, is the sum of its execution time and the completion time of the task ahead of it:
%\begin{equation} \label{eq:1}
%    C^{pend}_{ij} = E_{ij} + C_{(i-1)j} 
%\end{equation}
%\vspace{2pt}

\begin{figure}[h] 
  \centering
  \includegraphics[width=0.5\textwidth]{\paperfolder/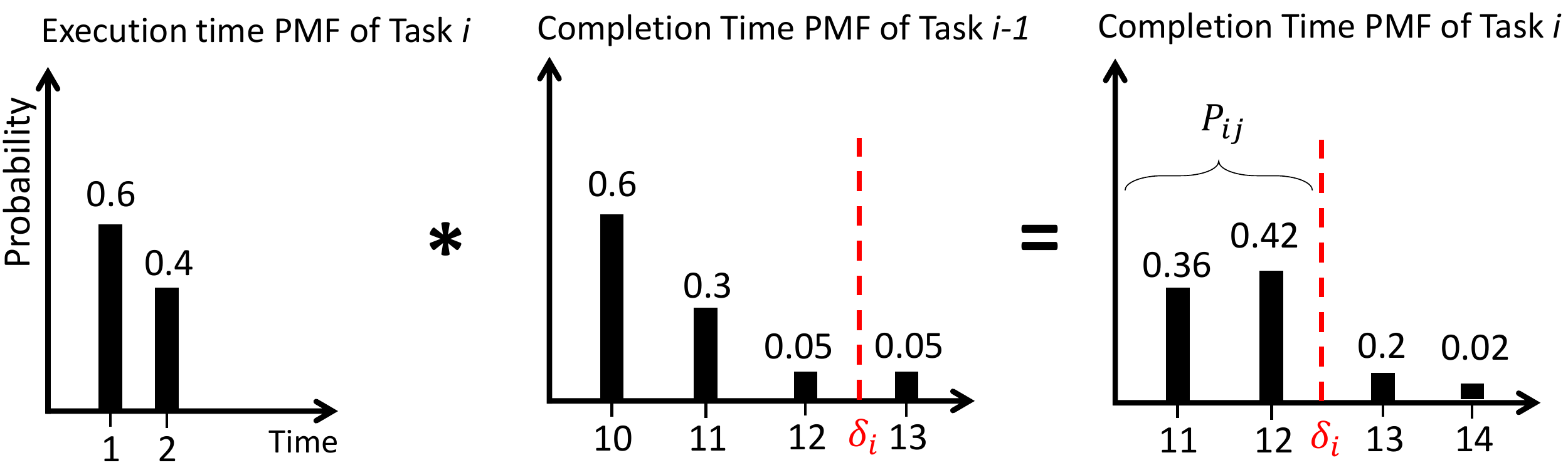}
  \caption{ Execution time PMF of pending task $i$ is convolved with the completion time PMF of task $i-1$ to obtain the completion time PMF of task $i$.\label{fig:convolution} }
\end{figure}

As depicted in Figure~\ref{fig:convolution}, to calculate the completion time PMF of pending task $i$ on the given machine $j$, its execution time PMF is convolved with the completion time PMF of the task ahead of it (\ie task $i-1$). Note that, if pending task $i$ cannot begin its execution before its deadline, denoted $\delta_i$, it is dropped. As this way of task dropping is performed in reaction to missing a task's deadline, we call it \emph{reactive task dropping}. Equation~\ref{eq:drop} shows the way $c_{ij}(t)$ is calculated. 
In this equation, if the completion time of task $i-1$ occurs at any time after $\delta_i$, task $i$ is reactively dropped, hence, its execution time is considered zero in the convolution process. In this case, $\forall t\geq \delta_i$, impulses of $c_{(i-1)j}(t)$ are directly added to $c_{ij}(t)$.
% Note that, in this equation, for $t\geq\delta_i$, task $i$ is reactively dropped, hence, its execution time is considered zero in the convolution process. That is, $\forall t\geq\delta_i$ completion time of task $i$ is the same as impulses in completion time of task $i-1$, therefore, these impulses ($c_{(i-1)j}(t)$ for $t\geq \delta_i$) are directly added to $c_{ij}(t)$.  

\begin{equation} \label{eq:drop}
    c_{ij}(t)= 
\begin{cases}
\sum\limits_{\forall k<t}  c_{(i-1)j}(k).e_{ij}(t-k), &  t < \delta_i\\
   \\
\sum\limits_{\forall k<\delta_i}  c_{(i-1)j}(k).e_{ij}(t-k) + c_{(i-1)j}(t), &  t \geq \delta_i
\end{cases}
\end{equation}

Once we calculate completion time PMF of task $i$, its chance of success, denoted $p_{ij}$, is calculated based on Equation~\ref{eq:3}.
\begin{equation} \label{eq:3}
    p_{ij} = \sum_{\forall t < \delta_i} c_{ij}(t)
\end{equation}

\begin{figure}[h] 
  \centering
  \includegraphics[width=0.4\textwidth]{\paperfolder/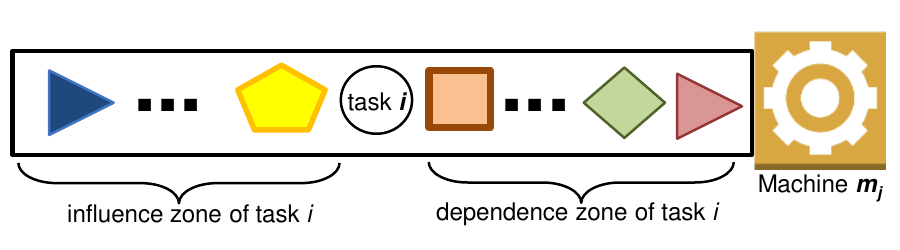}
  \caption{Stochastic completion time of task $i$ is dependent on the list of tasks ahead of it (dependence zone). Task $i$ influences stochastic completion time of tasks behind (influence zone).\label{fig:ZonesFig} }
  \vspace{-10px}
\end{figure}

Although task execution time is an independent random variable, task completion time is not. As depicted in Figure~\ref{fig:ZonesFig}, in a machine queue, completion time (and subsequently chance of success) of  task $i$ not only depends on its execution time, but also on the completion time of the tasks ahead of it, defined as \emph{dependence zone}. Similarly, task $i$ influences the completion time of the tasks behind it in the machine queue, defined as \emph{influence zone}. Note that, upon dropping task $i$, only its influence zone is affected.

\subsection{Calculating Chance of Success in Proactive Task Dropping} \label{sec:DropDecision}
In this part, we investigate how predictively deciding to drop a task, known as \emph{proactive task dropping}, can favor the overall system robustness. For that purpose, we need to measure the potential benefit of task dropping on the system robustness. For a list of $q$ pending tasks in machine queue $j$, we define \emph{instantaneous robustness}, denoted $R_j$, as the sum of their chances of success and calculate it based on Equation~\ref{eq:irobust}. Our hypothesis is that the overall system robustness is likely to be improved, only if instantaneous robustness is improved at each individual mapping event.

\begin{equation} \label{eq:irobust}
    R_j = \sum_{i = 0}^{q} p_{ij}
\end{equation}

Because dropping task $i$ only affects the chance of success for tasks in its influence zone, dropping task $i$ is considered appropriate, only if it improves the instantaneous robustness of tasks in the influence zone. For task $i$ in the machine queue, we need a method to calculate the instantaneous robustness of its influence zone in two cases: when task $i$ is not dropped versus when it is provisionally dropped. 

%To measure the effect of dropping task $i$, firstly, it is assumed that task $i$ is dropped from machine queue $j$. 
Upon provisional dropping of task $i$, the completion time of task $i-1$ is convolved with the execution time of task $i+1$ with respect to its deadline ($\delta_{i+1}$), as explained in Equation~\ref{eq:drop}. Let $c^{(i)}_{(i+1)j}(t)$ represent the completion time PMF of task $i+1$ when task $i$ is provisionally dropped. Formally, $c^{(i)}_{(i+1)j}(t)$ is calculated based on Equation~\ref{eq:dropcomp}.

 \begin{equation} \label{eq:dropcomp}
    c^{(i)}_{(i+1)j}(t)= 
\begin{cases}
\sum\limits_{k=0}^{k<t}  c_{(i-1)j}(k).e_{(i+1)j}(t-k), &  t < \delta_{(i+1)}\\
   \\
\sum\limits_{k=0}^{k<\delta_{i+1}}  c_{(i-1)j}(k).e_{(i+1)j}(t-k)
\\
                    + c_{(i-1)j}(t), &  t \geq \delta_{(i+1)}
 \end{cases}
\end{equation}

Accordingly, completion time PMF of next tasks in the influence zone of task $i$, $c^{(i)}_{nj}(t)$ for $\forall n\ge (i+2)$, is determined using Equation~\ref{eq:5}.

 \begin{equation} \label{eq:5}
    c^{(i)}_{nj}(t)= 
\begin{cases}
\sum\limits_{k=0}^{k<t}  c^{(i)}_{(n-1)j}(k).e_{nj}(t-k), &  t < \delta_n\\
   \\
\sum\limits_{k=0}^{k<\delta_n}  c^{(i)}_{(n-1)j}(k).e_{nj}(t-k) + c^{(i)}_{(n-1)j}(t), &  t \geq \delta_n
\end{cases}
\end{equation}

Once we have the completion time PMF for task $n$ in the influence zone, its chance of success, denoted $ p^{(i)}_{nj}$, is calculated based on Equation~\ref{eq:6}.
\begin{equation} \label{eq:6}
    p^{(i)}_{nj} = \sum_{t = 0}^{ t < \delta_n} c^{(i)}_{nj}(t)
\end{equation}

\subsection{Optimal Proactive Task Dropping} \label{sec:OptimalDropping}
Recall that dropping a task has two contradictory effects on the system robustness. Although it reduces the number of completed tasks by one, it increases the chance of success in its influence zone, and therefore, the instantaneous robustness. 

Due to the impact of proactive task dropping on the chance of success of tasks in its influence zone, proactive dropping is not an independent decision to be made for a task in isolation. As an example, assume that task $n$ is located in the influence zone of a large (\ie compute intensive) task $i$, such that the chance of success for $n$ tends to zero ($p_{nj}\to 0$). Therefore, instantaneous robustness does not gain from dropping $n$. However, proactively dropping $i$ can affect the chance of success for $n$ and make it appropriate for dropping. We can conclude that an optimal proactive task dropping must maintain a collective view to the list of tasks of a machine queue, as opposed to deciding for each task in isolation. Thus, the problem of optimal proactive dropping is narrowed down to finding a subset of tasks whose dropping maximizes the instantaneous robustness. 

As the influence zone of the last task in a machine queue is null, its dropping does not improve instantaneous robustness, hence, it is excluded from the subset of tasks considered for proactive dropping decision. Accordingly, in a machine with queue size $q$, finding the optimal proactive dropping decision requires $2^{q-1}$ subsets to be examined for dropping. The subset of tasks whose dropping maximizes the instantaneous robustness represents the optimal proactive dropping decision.

\subsection{Proactive Task Dropping Heuristic}\label{subsec:heurdrop}
As finding the optimal subset of tasks for proactive dropping includes examining an exponential number of cases, it imposes a considerable overhead at each mapping event. As such, in this part, we propose a task dropping heuristic that provides a sub-optimal solution within a feasible time. The proposed heuristic does not examine all subsets, instead, it operates on a task by task basis and decides about the proactive dropping of each task. Specifically, the heuristic iterates each machine queue and only in one pass decides appropriate tasks for proactive dropping. 
%TODO: can we say how many of subsets at max are examined?

The appropriateness of proactively dropping task $i$ can be measured by comparing the instantaneous robustness of machine $j$ when task $i$ is provisionally dropped, denoted $R^{(i)}_j$, versus the circumstance in which task $i$ is not dropped (\ie $R_j$). Let $Q_j$ represent the list of pending tasks on machine queue $j$, then $R^{(i)}_j$ is calculated based on Equation~\ref{eq:7}.

\begin{equation} \label{eq:7}
    R^{(i)}_j = \sum_{\forall n\in Q_j - \{i\}} p^{(i)}_{nj}
\end{equation}

In particular, dropping task $i$ is considered appropriate, if $ R^{(i)}_j$ is sufficiently greater than $R_j$. That is, we should have $R^{(i)}_j > \beta\cdotp R_j$, where $\beta\geq 1$ is defined as the \emph{robustness improvement factor}. In fact, the value of $\beta$ dictates the aggression level of proactive task dropping. In spectrum, $\beta\to \infty$ disables proactive task dropping whereas
$\beta\to 1$ enacts dropping even for minor improvements in instantaneous robustness. We study the suitable value for $\beta$ in the evaluation section of the paper.

Note that, in examining provisional dropping of task $i$, only its influence zone has to be considered and the dependence zone of the task can be excluded for the calculations. We argue that, there is not much gain in exploring the whole influence zone. Knowing that provisional dropping of task $i$ decreases the instantaneous robustness by $p_{ij}$. Assuming $\beta=1$, the gain in the instantaneous robustness of tasks in the influence zone must be greater than $p_{ij}$, so that proactive dropping of task $i$ is enacted. Theoretically, the gain can occur due to accumulation of negligible improvements across a large number of tasks that eventually may not increase the system robustness. To avoid dropping because of such misleading gains, in this heuristic, we enact proactive dropping of task $i$, if the loss in the instantaneous robustness is compensated only within the first few tasks of the influence zone. 

For proactive task dropping heuristic, we define \emph{effective depth}, denoted $\eta$, as the number of tasks located immediately after task $i$ in its influence zone. Then, robustness improvement is only examined for tasks $n\in <i+1,...,i+\eta>$. In summary, proactive dropping of task $i$ on machine $j$ is confirmed by the heuristic, only if the condition in Equation~\ref{eq:1} holds.

\begin{equation}\label{eq:1}
   R^{(i)}_j > \beta\cdotp R_j  \iff \sum_{n = i+1}^{i+\eta} p^{(i)}_{nj} > \beta\cdotp\sum_{n = i}^{i+\eta} p_{nj}
\end{equation}

The algorithm in Figure~\ref{fig:algdrop} explains the proactive task dropping heuristic. In the first step, the algorithm iterates through all machine queues and performs reactive task dropping for those already missed their deadlines (as noted in Step 2). Then, in Steps 4---9, for each task $i$, we examine provisionally dropping it and compare the instantaneous robustness for the effective depth of task $i$ with the circumstance that task $i$ is not dropped. In Step 9, task $i$ is proactively dropped if the condition in Equation~\ref{eq:1} holds.
Mapping heuristic is invoked after the proactive dropping heuristic.

\boxfig{

$\beta \leftarrow$ Robustness Improvement Factor

$\eta \leftarrow$ Effective Depth
%$c \leftarrow$  Fairness Factor

%$\gamma_{1..n} \leftarrow$ Fairness Score of all task types
%$\beta_r$ = $DroppingThreshold$

%$\beta_e$ = $DeferringThreshold$

\bigbreak

Upon triggering of a mapping event: 
\begin{itemize}
\item[(1)] For each queue $j$ of machines $\{m_0,m_1,...,m_s\}$: 
\begin{itemize}
\item[(2)] Drop all pending tasks that missed their deadlines
%\item[(3)] Collect and process data of all task $t_f$ completed on-time since previous mapping event
% \begin{itemize}
% 	\item[--] $k \leftarrow $ task type of $t_f$ 	
%	\item[--] $\gamma_k \leftarrow \gamma_k - c$
% \end{itemize}  
\item[(3)] For each task $i$ in machine queue $j$:
		\begin{itemize}
		    \item[(4)] For each task $n$ in effective depth of task $i$:
		        \begin{itemize}
		            \item[(5)] Calculate $p_{nj}$ based on Equation~\ref{eq:3}
		            \item[(6)] Provisionally drop task $i$ 
		            \item[(7)] Calculate $p^{(i)}_{nj}$ based on Equation~\ref{eq:6}
		        \end{itemize}
		    \item[(8)] if $\sum_{n=i+1}^{i+\eta} p^{(i)}_{nj} > \beta \cdotp \sum_{n=i}^{i+\eta} p^{(i)}_{nj} $
		          \begin{itemize}
		                \item[(9)] Confirm dropping of task $i$
		           \end{itemize}
		 \end{itemize}
\end{itemize}

	\item[(10)] Call mapping heuristic
% 	\item[(10)] For each task $t_i$  of type $k$ mapped to machine $j$:
% 	\begin{itemize}

% 		\item[(11)] Dispatch remaining assigned tasks to machines
% 		\
% 	\end{itemize}
\end{itemize}

\caption{\small{Pseudo-code for Proactive Task Dropping Heuristic.}}
 \label{fig:algdrop}
}

%%%%%%%%new for CCGrid submission
\subsection{Complexity Analysis of Proactive Task Dropping}
\label{subsec:complexity}
Time complexity of proactive task dropping in each mapping event depends on two factors: (A) the number of convolutions; and (B) the complexity of performing each convolution.

As noted earlier, the number of cases that the optimal dropping examines is $2^{q-1}$ and for each case, at most $q$ tasks are considered. Hence, in the worst case, the number of convolutions required (factor A) for the optimal solution is $O(q\cdotp2^{q-1})$. Alternatively, heuristic dropping approximates optimal dropping by iterating the machine queue from the head to the tail only once, evaluating the impact of dropping each task on $\eta$ tasks in its influence zone. Therefore, it requires at most \textbf{$O(\eta \cdotp q$)} convolutions. 

Let $N_1$ and $N_2$ the set of impulses of two given PMFs. In the worst case, we assume that the PMFs are such that the number of impulses in the convolved PMF is $|N_1|\cdotp|N_2|$. Then, the time complexity of the convolution operation (factor B) is $O(N^2$), where $N=max(|N_1|,|N_2|)$. Accordingly, calculating the completion time of all tasks in a machine queue with size $q$ has a time complexity of $O(N^q)$ where $N=max_{i=1}^{q} (|N_i|)$. As a result, the overall time complexity of the proactive task dropping heuristic is $O(q\cdotp N^q)$. Note that, in practice, the value of $q$ is low and, based on our observations, the number of impulses generated by a convolution is far less than $|N_1|\cdotp|N_2|$.% Also, the number of impulses is independent of the task execution time.
% A long executing task should have impulses with lower granularity (\ie each impulse represent a long time-span) than a shorter task.
%A long executing task should not have many more impulse count than a shorter executing task. 

%, Instantaneous robustness information is calculated as the summation 
%To obtain instantaneous robustness data of machine queue $Q_j$, we simply calculate the summation of all task's chance of success, which can be obtained in linear time complexity of the number of tasks (let this be $N_j$). In the optimal 

%While time complexity of heuristic-based proactive task dropping is higher than the threshold-based probabilistic task dropping ($q$ convolutions), the out-performance of its dropping decisions can allow the system to save time on the scheduling side. Especially if simpler mapping heuristic can be used in place of a complex mapping heuristics with probabilistic task deferring. %If the machine queue length is limited and the number of unmapped task is higher than $2 \cdotp \eta$, then extra dropping decision making overhead is offset by skipping ...
%==============
%\input{Sources/DropDecision}

\section{Performance Evaluation and Analysis}
\label{sec:evltn}

\subsection{Experimental Setup}
To evaluate the task dropping mechanism, we simulate two scenarios: one using four video transcoding as task types and four AWS cloud virtual macine (VM) as the HC system. To study the mechanism further, we simulate a more diverse HC system with eight machines and twelve task types from  SPECint \cite{specint} benchmarks. We base our analysis on the latter workload because it provides a wide variety of inconsistent heterogeneous workload. Then, we use the cloud-base workload for validation of the findings.

The eight machines in the latter scenario contains eight machines\footnote{The 8 machines are: Dell Precision 380 3 GHz Pentium Extreme, Apple iMac 2 GHz Intel Core Duo, Apple XServe 2 GHz Intel Core Duo, IBM System X 3455 AMD
Opteron 2347, Shuttle SN25P AMD Athlon 64 FX-60, IBM System P 570 4.7 GHz,
SunFire 3800, and IBM BladeCenter HS21XM.}. 
The function describing execution time of the tasks on a machine is assumed to be a unimodal distribution. Gamma distribution was used to generate the distributions and the mean of the Gamma distribution was determined based on execution time results of SPECint benchmarks on the aforementioned eight machines. We sampled 500 execution times for each application on each machine where the scale parameter of each Gamma distribution was chosen uniformly from the range [1,20]. Once the sample execution times were generated, we applied a histogram to discretize the result and produce PMFs. The PMFs of different benchmarks on the eight heterogeneous machines collectively form the PET matrix.

PET matrix of the eight machines by twelve task types are used throughout the experiments. Each machine is provided with a machine-queue which can store up to six tasks, including the task that is currently executing. Task dropping mechanism is engaged each time a system notices a task missing its deadline. All the experiments are performed on Louisiana Optical Network Infrastructure (LONI) Queen Bee 2 HPC system~\cite{LONI}.

Each simulation starts and ends when the system is in the idle state. In a simulation, each task arrives based on an arrival time and eventually oversubscribe the system. As we focus on the oversubscribed condition, the first and last 100 tasks in each workload trial are excluded from the results. For each experiment, 30 workload trials with the same intensity level were examined. Workload intensity refers to the number of tasks per time unit arrive to the system. For each experimental result, the mean and 95\% confidence interval are reported.

Every workload trial introduces a level of oversubscription to the system, such that all tasks cannot complete successfully, due to shortage of the resources. However, every single task is individually feasible to process on time. Deadline for any given task $i$ is determined based on  $\delta_i = arr_i + avg_i + (\gamma \cdotp avg_{all})$, where $arr_i$ is the arrival time, $avg_i$ is the mean execution time for the task type (range from 50 to 200 ms), $\gamma$ is a coefficient determining the task slack, and $avg_{all}$ is the mean of all task types execution times. To evaluate the system robustness against task arrival uncertainty, we conduct all experiments with three levels of task arrival intensity, creating workloads with 20K, 30K, and 40K tasks.

\subsection{Mapping Heuristics}\label{subsec:baseline}
The dropping mechanism introduced in this paper is generic and independent from any particular mapping heuristic. In fact, dropping mechanism can be considered as a separate component in a resource allocation system that can cooperate with any mapping heuristic, such as those widely-used in heterogeneous systems (\eg MinMin \cite{salehi2016stochastic}, MSD \cite{TPDS17}, and PAM \cite{ipdps19}) or homogeneous systems (\eg FCFS, SJF, and EDF), to improve the system robustness. 

% For each heuristic, we measure the system robustness resulted from the cooperation of that mapping heuristic with proactive task dropping. We are also interested to see how much the system robustness is improved when PAM cooperates with the proposed proactive dropping mechanism versus when PAM is employed with its original threshold-based dropping settings. 

\subsubsection{MinCompletion-MinCompletion (MinMin or  MM)}
MinMin (MM) is a popular mapping heuristic in heterogeneous computing literature~\cite{pedemonte2016accelerating, salehi2016stochastic}. %cite removed{he2003qos,ezzatti2013efficient}
In the first phase of this heuristic, for each task in the batch queue, the machine that offers the minimum expected completion time is found, and a pair is formed. In the second phase, for each machine with an available slot in its queue, from the task-machine pairs provisionally mapped to that machine, the pair with the  minimum completion time is assigned to it. The process is repeated until all machine queues are full, or until the batch queue is depleted.

\subsubsection{MinCompletion-Soonest Deadline (MSD)}
Similar to MinMin, MSD is also a two-phase mapping heuristic used in several earlier sutdies (\eg \cite{khemka2015utility,salehi2016stochastic,ipdps19}). The first phase creates task-machine pairs based on minimum expected completion time for each unmapped task. In the second phase, for each machine with a free slot, the task-machine pair that has the soonest deadline is assigned to that machine. 

Ties are broken by choosing the task that has the minimum expected completion time. Similar to MM, after assigning tasks to free slots, the operation is repeated until either there is no unmapped task or there is no free slot in machine queues.

\subsubsection{Pruning-Aware Mapping (PAM)}
PAM \cite{ipdps19} is a state of the art heuristic functions based on the PET matrix and operates based on the chance of success for tasks. PAM is a two-phase mapping heuristic. In its first phase, for each task, it finds the machine provides the highest chance of success. Then, the second phase finds the task-machine pair with the lowest completion time and maps it to that machine queue. Ties are broken by assigning the task that has the shortest expected execution time. PAM performs task dropping (from machine queues) and task deferring (from the batch queue) at each mapping event. However, because this study focuses on the dropping operation, for the sake of comparison, we disabled deferring on PAM. 

PAM uses a predetermined threshold for dropping and deferring decisions. We replace the dropping thresholds of PAM with our proposed proactive dropping mechanism. Specifically, we consider two separate cases for evaluation: (A) Combination of PAM with optimal proactive task dropping (shown as $PAM+Optimal$); (B) Combination of PAM with heuristic proactive task dropping (shown as $PAM+Heuristic$).

%========= below are experiments

% \subsection{Impact of Deferring Threshold Value Evaluation}
% In this experiment, our goal is to study the usefulness of  operating deferring mechanism and dropping decision together. To this end, we test the system using the deferring threshold range from 0\% (\ie no defer) to 95\%. Recall that deferring threshold is the factor used to postpone tasks with chance of success lower than deferring threshold. Then, the deferred task may have a higher chance of success at the next mapping events by waiting for a machine with better affinity with the task type.

% \begin{figure} 
%   \centering
%   \includegraphics[width=0.5\textwidth]{\paperfolder/Figures/R_Defer_30k.pdf}
%   \caption{The impact Deferring on the system robustness with Proactive Dropping. \label{R_Defer} }
% \end{figure}

% Figure \ref{R_Defer} shows that the best performance obtained when deferring threshold is set to 0\%. The robustness is decreased by applying higher values of deferring threshold to the system until it reaches the absolute minimum point at deferring threshold of 60\%. Further increasing the deferring threshold leads to increasing the robustness. At deferring threshold of 90\%, the system performance begins degrading again. 
% In the rest of the experiments, the dropping mechanism is used without any deferring.  

\subsection{Analyzing the Impact of Effective Depth}
\begin{figure} 
  \centering
  \includegraphics[width=0.36\textwidth]{\paperfolder/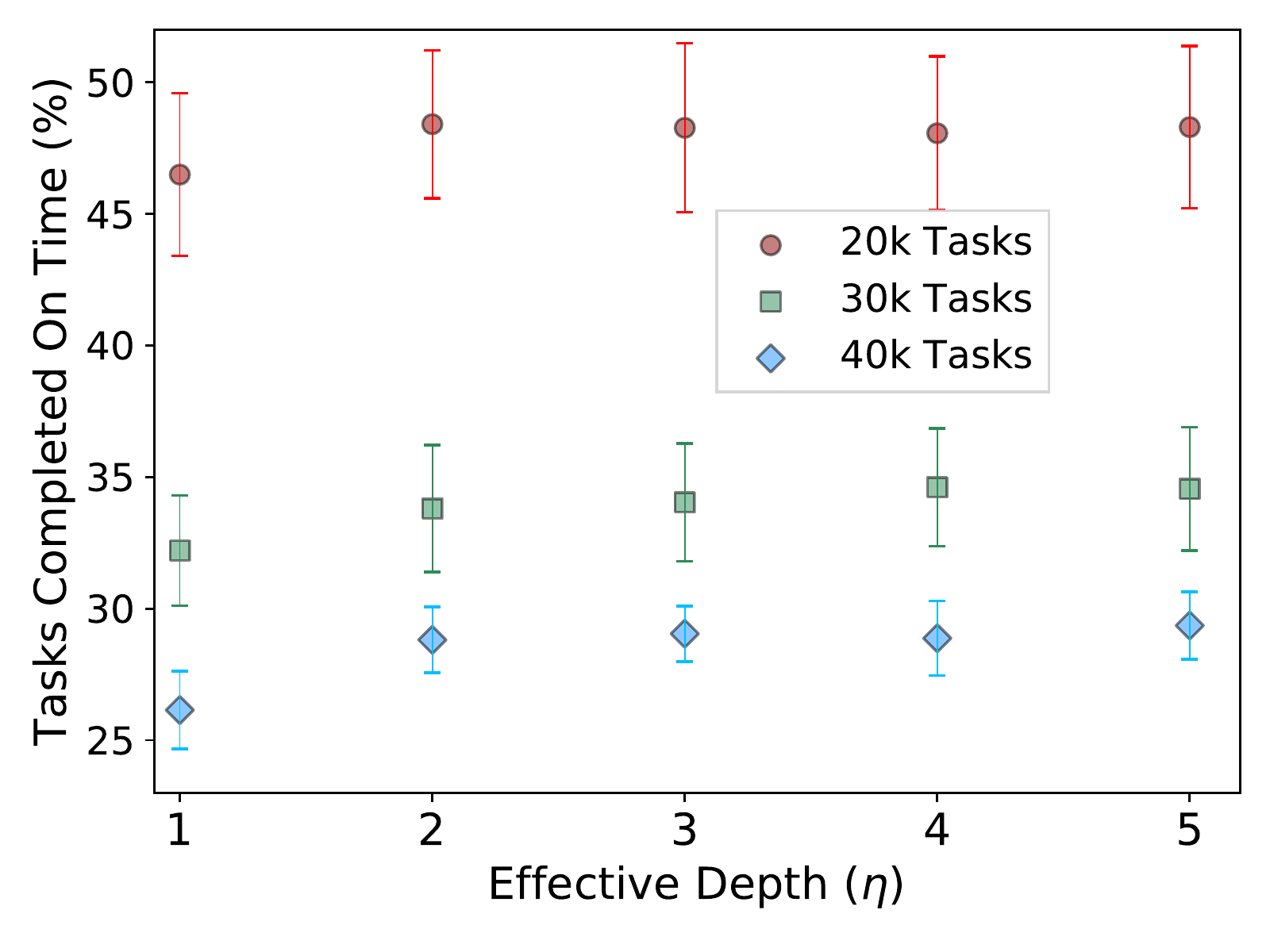}
  \caption{The impact of varying effective depth on the system robustness resulted from proactive task dropping heuristic with PAM mapping heuristic. The horizontal axis shows the effective depth ($\eta$) and vertical axis shows the system robustness in form of the percentage of tasks completed on time. \label{fig:R_eta} }
\vspace{-5px}
\end{figure}

In Section~\ref{subsec:heurdrop}, we described that proactive task dropping heuristic does not need to examine the whole influence zone of a task to decide about its dropping. In this part, we aim at identifying the suitable number of tasks in the influence zone (\ie effective depth), whose robustness improvement should compensate for the loss of robustness resulted from a task dropping. For that purpose, we analyze how the robustness of an HC system differs by varying the values of effective depth ($\eta$). The result of this analysis is shown in Figure \ref{fig:R_eta}. The horizontal axis shows different values of effective depth and the vertical axis shows the system robustness in form of percentage of tasks completed on time. The experiment was conducted for three oversubscription levels.

As shown in Figure~\ref{fig:R_eta}, there is no significant improvement in the system robustness for $\eta>2$. The reasons are twofold: First, considering too many tasks for effective depth can be misleading to the task dropping heuristic. This is because the robustness loss resulted from dropping task $i$ can be potentially amortized across multiple tasks in the influence zone, causing a slight (but practically ineffective) improvement in their chances of success. In this circumstance, the task dropping heuristic malfunctions by suggesting dropping task $i$, without necessarily improving the number of tasks completing on time. This observation confirms our hypothesis in Section~\ref{subsec:heurdrop}. Second, from a probabilistic point of view, when we drop task $i$ in an oversubscribed system, the uncertainty exists in the completion time of tasks located immediately after task $i$ gradually absorb the gain of robustness resulted from dropping task $i$. We can conclude that, in an oversubscribed system, the impact of dropping task $i$ fades out quickly, within the first couple of tasks in the influence zone of task $i$. %TODO: how about checking lower oversubscription? How about more slack? How about checking gradual fade of in instant robustness as we move forward in the influence zone

Although the above justification suggests effective depth to be small, in Figure~\ref{fig:R_eta}, we  observe that effective depth of 1 is not effective. In fact, the case of $\eta=1$ can be misleading in certain circumstances. For example, consider task $i$ is unlikely to succeed (say $p_{ij}=10\%$), therefore, it is provisionally dropped. However, task $i+1$ is already likely to succeed (say $p_{ij}=95\%$) and provisionally dropping $i$ can improve chance of task $i+1$ by at most $5\%$. Because the robustness improvement cannot compensate the loss of it (which is 10\% by dropping task $i$), dropping heuristic decides not to drop task $i$. However, because  $\eta=1$, the heuristic neglects considering task $i+2$ in the influence zone that can potentially gain significantly from dropping task $i$. According to this analysis, for the rest of evaluations, we configure the proactive mapping heuristic to be carried out with $\eta = 2$.  

\subsection{Analyzing the Impact of Robustness Improvement Factor}
\begin{figure} 
  \centering
  \includegraphics[width=0.37\textwidth]{\paperfolder/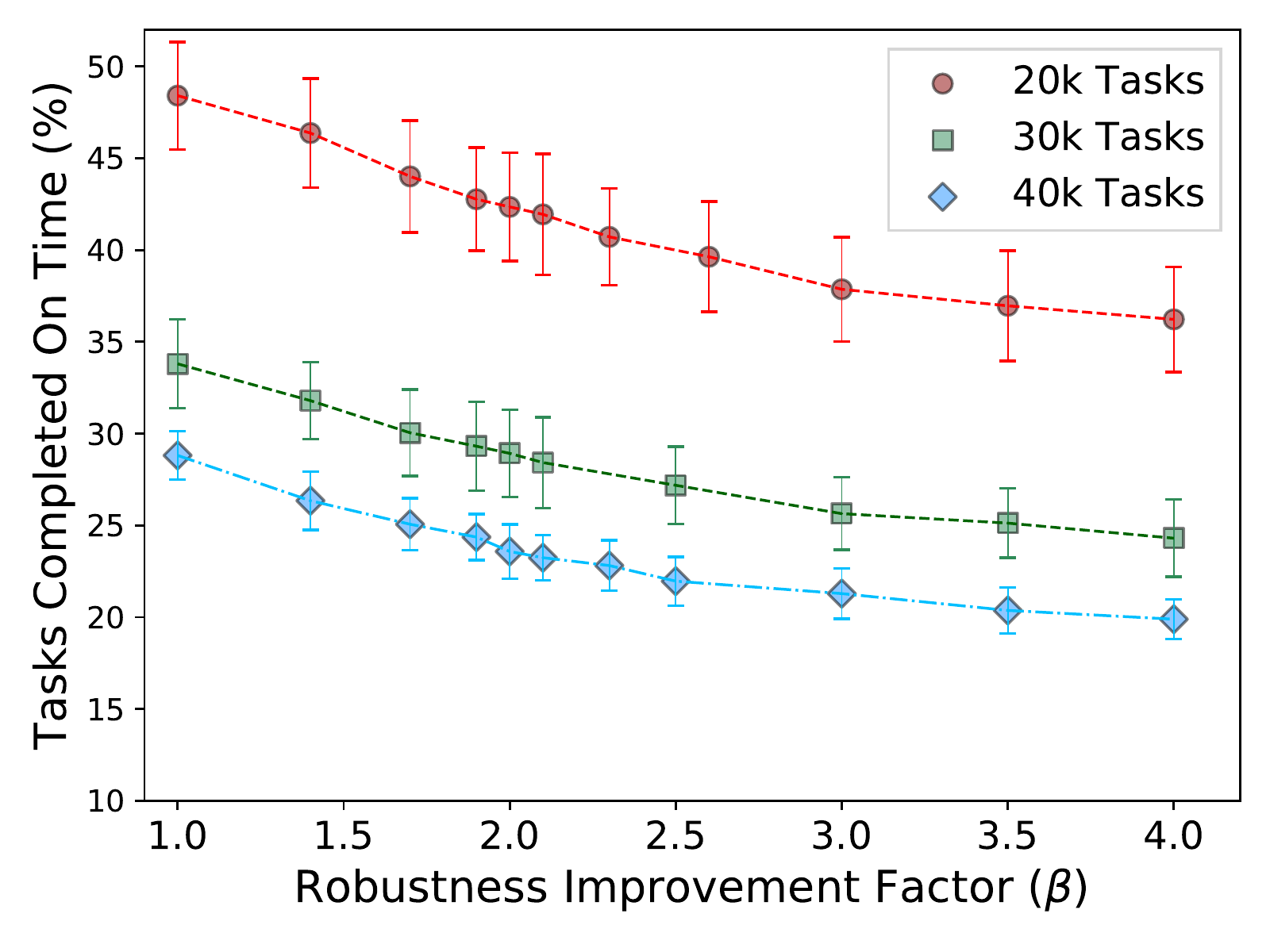}
  \caption{The impact of Robustness Improvement Factor ($\beta$ in horizontal axis) on the system robustness resulted from proactive task dropping heuristic with PAM mapping heuristic for different oversubscription levels. \label{R_beta} }
\end{figure}

As we described in Section~\ref{subsec:heurdrop}, the proactive task dropping heuristic decides about appropriateness of a task dropping based on a Robustness Improvement Factor ($\beta$). In this part, we experimentally identify the suitable value that should be considered for $\beta$, so that the system robustness gain is maximized. To this end, as shown in Figure \ref{R_beta}, we vary the value of $\beta$ in the range of [1,4] by step 0.5 and, for each configuration, we measure the system robustness in form of percentage of tasks completed on time. We conduct the experiment for all three levels of oversubscription. %We note that, we conducted the further analysis on other po 

\begin{figure*}[h]
  \begin{subfigure}{0.5\textwidth}
     \centering
     \includegraphics[width=0.72\linewidth]{\paperfolder/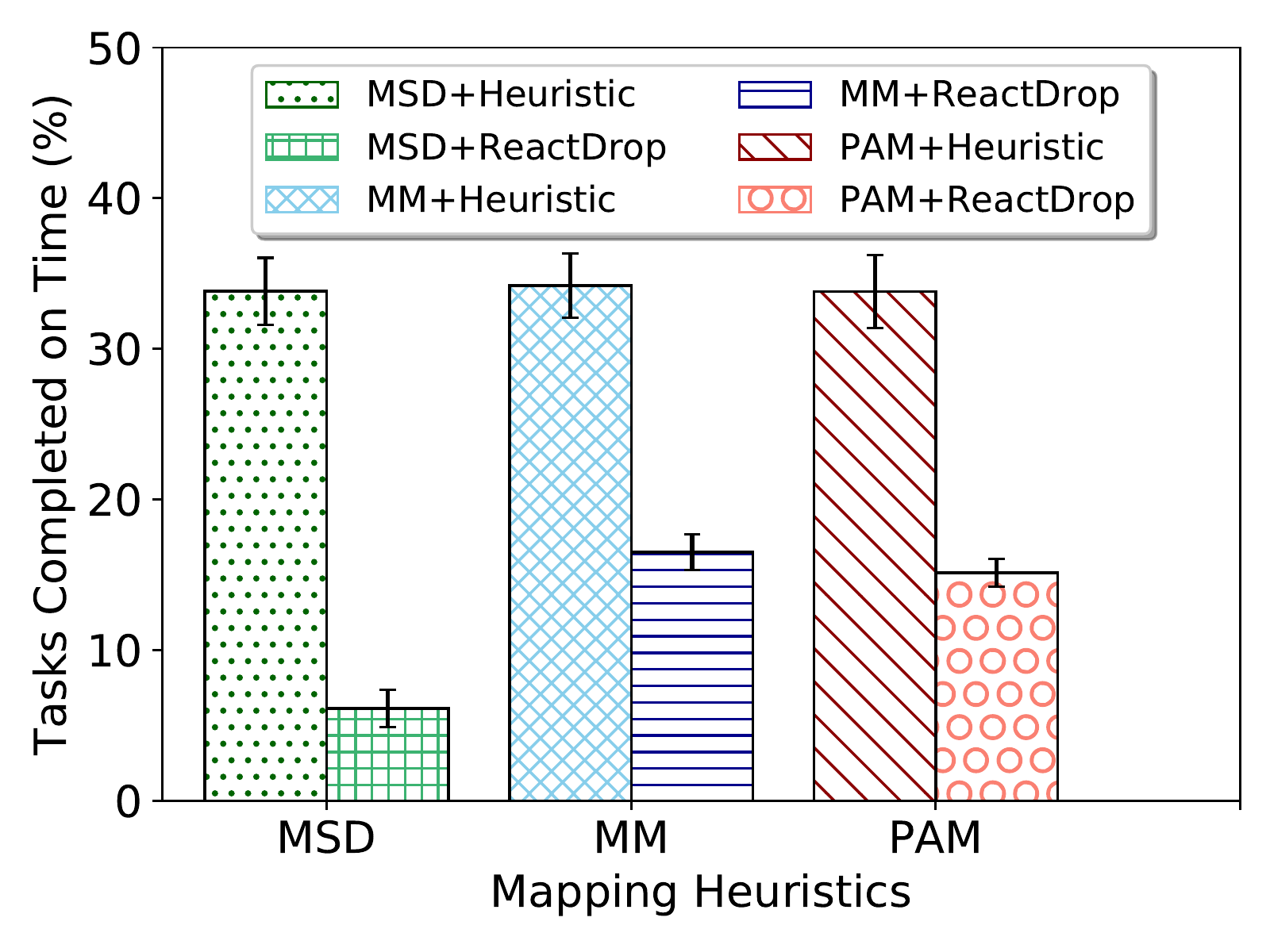}
  
     \caption{Proactive task dropping in a heterogeneous system}
     \label{subfig:dropheur}
  \end{subfigure}\hfill
  \begin{subfigure}{0.5\textwidth}
     \centering
     \includegraphics[width=0.72\linewidth]{\paperfolder/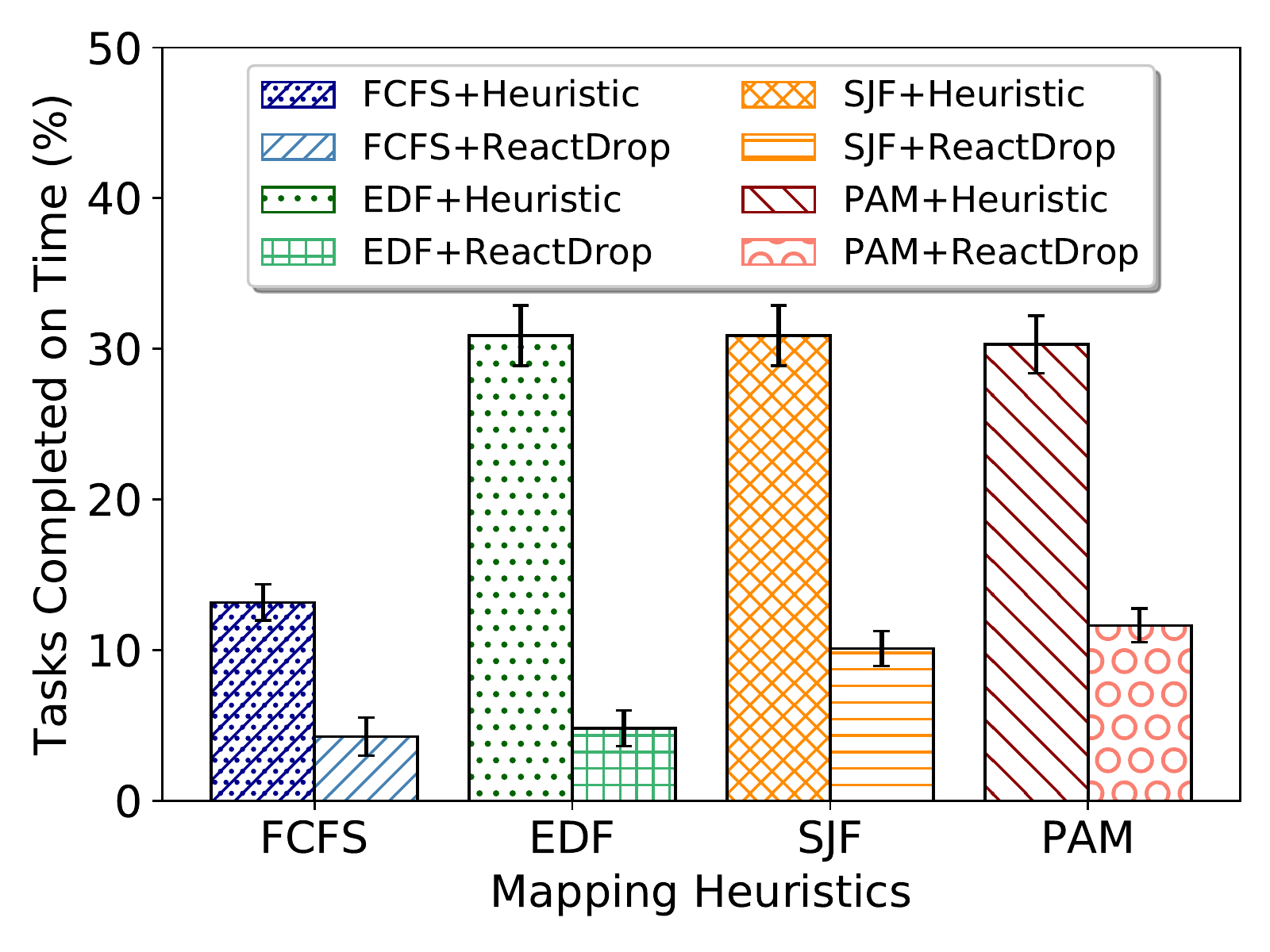}
     \caption{Proactive task dropping in a homogeneous system}
   \label{subfig:dropheurhomo}
  \end{subfigure}\hfill
  \caption{Evaluating the impact of applying proactive task dropping heuristic to different mapping heuristics. Subfigure (a) shows the results for a heterogeneous computing system and Subfigure (b) shows it in a homogeneous computing system. Horizontal axes show different mapping heuristics---each one deployed with proactive task dropping heuristic (+Heuristic) and without proactive task dropping heuristic (+ReactDrop). In each case, system robustness in form of percentage of tasks completing on time is reported.}
      \vspace{-3mm}

  \label{fig:combination}
\end{figure*}

As we can see in this figure, the system robustness is maximized for $\beta = 1$ and the system robustness declines, as the $\beta$ value increases. In fact, by increasing the $\beta$ value proactive task dropping heuristic becomes more conservative and is less often engaged in the task dropping operation. At the end of the spectrum, very large values for $\beta$ neutralizes the impact of proactive dropping heuristic. According to this analysis, for the rest of evaluations, we configure the proactive mapping heuristic to be carried out with $\beta = 1$.

\subsection{Analyzing the Impact of Proactive Task Dropping on Various Mapping Heuristics}\label{subsec:allheur}

Although the proposed task dropping mechanism is independent from mapping heuristics, the two can have a synergy in achieving robustness against the compound uncertainty. To examine both the generality of the dropping mechanism and its impact on the system robustness, in this experiment, we apply the proactive task dropping heuristic on widely-used mapping heuristics of both heterogeneous and homogeneous systems. Then, for each mapping heuristic, we measure the system robustness (percentage of tasks completing on time) with proactive task dropping heuristic (+Heuristic) and without proactive task dropping heuristic involved (+ReactDrop). In this experiment, the oversubscription level of the system is set on 30K tasks. 

The results of this experiment are shown in Figure~\ref{fig:combination}. Subfigure \ref{subfig:dropheur} shows the percentage of tasks completed on time (vertical axis) and its horizontal axis shows MSD, MM, and PAM mapping heuristics, each one with and without proactive task dropping heuristic. In this figure, we observe that when proactive task dropping is not applied, MSD performs significantly lower than MM and PAM. This is because in an oversubscribed system, mapping tasks based on their deadline intensity implies allocating tasks with a low chance of success and postponing tasks that have a high chance of success to a later time. However, we observe that when proactive task dropping is in place, all three mapping heuristics provide almost the same robustness. This is because proactive task dropping prunes tasks whose chance of success is low from machine queues. Interestingly, the results show that, if we put a reasonable dropping mechanism in place, we do not have to deploy a complex mapping heuristic. In this case, simple mapping heuristics can be forgiven for their poor mapping decisions and ultimately provide a competitive robustness.

The result of this experiment for homogeneous mapping heuristics is shown in Figure~\ref{subfig:dropheurhomo}. In this experiment, we employed three mapping heuristics that are popular in homogeneous systems, namely FCFS, SJF, and EDF (earliest deadline first) and a prior work's mapping heuristic named PAM. The figure testifies that the dropping mechanism can significantly improve the robustness of homogeneous systems. We observe that, without dropping, FCFS and EDF provide the lowest robustness. The reason that SJF and PAM provide better robustness is that SJF always maps the shortest tasks, hence, can increase the number of completed tasks. Also, PAM always maps the ones with the highest chance that leads to completing tasks on time. Similar to heterogeneous systems, we observe that proactive dropping heuristic can compensate poor decisions made by mapping heuristics and increase their robustness to almost the same magnitude. The improvement in robustness is less significant for FCFS. This is because, unlike SJF, in FCFS, executing a compute-intensive task can diminish the chance of success for several pending tasks, such that even by proactively dropping them the chance of success for remaining tasks does not improve significantly. 

\subsection{Analyzing the Impact of Proactive Task Dropping on the System Robustness}
In this experiment, our goal is to evaluate how proactive dropping can enhance the system robustness against compound uncertainty in both task execution times and arrival. 
Based on the previous experiment, we pick PAM as the mapping heuristic for this study and apply the following four variations of task dropping on it: (A) using optimal proactive dropping (termed PAM+Optimal); (B) using proactive dropping heuristic (termed PAM+Heuristic); and (C) using a threshold based approach (termed PAM+Threshold). Case (C) was developed in \cite{ipdps19} and in that the system user needs to be aware of dropping and initially set its threshold. Then, the predetermined threshold is adjusted at each mapping event.  

\begin{figure} [hb]
  \centering
  \includegraphics[width=0.35\textwidth]{\paperfolder/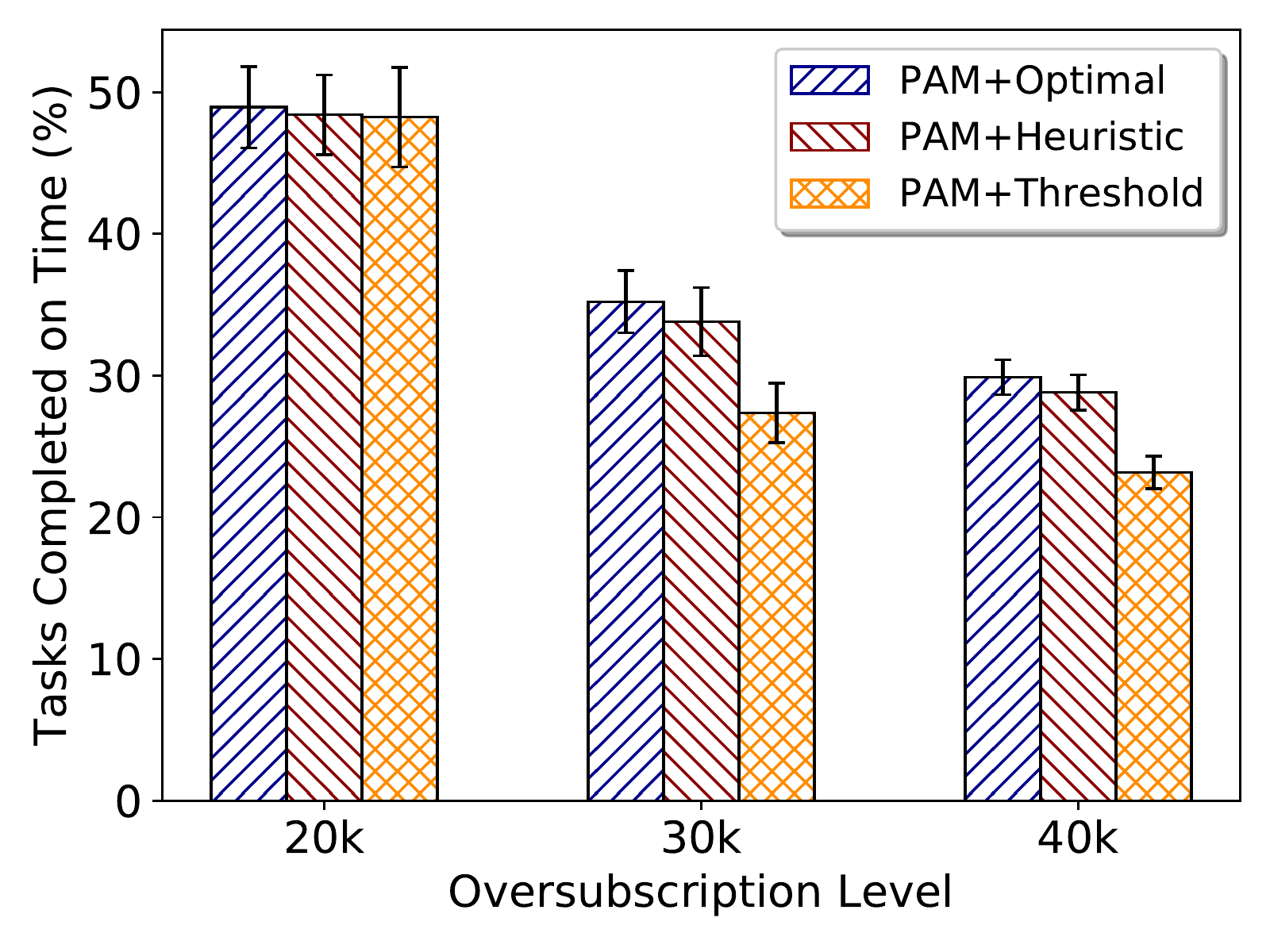}
  \caption{Comparing the impact of proactive task dropping against other forms of task dropping in terms of system robustness, measured by the percentage of tasks completed on time (vertical axis). The experiment is conducted for various oversubscription levels (horizontal axis). \label{fig:dropping} }
\end{figure}

Figure~\ref{fig:dropping} shows the result of evaluating variations of task dropping across three oversubscription levels, represented by the number of arriving tasks (as shown in the horizontal axis). In each case, we measure the system robustness in form of percentage of tasks completing on time (vertical axis).

The experiment results demonstrate that as the systems becomes more oversubscribed, the system robustness declines. However, we observe that both PAM+Optimal and PAM+Heuristic outperform PAM+Threshold. Specifically, when the system is under 40K task arrival, both PAM+Optimal and PAM+Heuristic outperform PAM+Threshold by around 8\%. The results indicate the efficacy of the proactive dropping approaches. This improvement is particularly remarkable, when considering that proactive dropping is also less complicated than PAM+Threshold and it does not require any user involvement in adjusting dropping threshold.

Further analysis between PAM+Optimal and PAM+Heuristic reveal that, regardless of the oversubscription level, there is no statistically and practically significant difference between these two approaches. Considering simplicity and competitive performance of PAM+heuristic, we can conclude that it can replace PAM+Optimal without any major loss in robustness. 

To analyze the impact of proactive task dropping on the observed robustness, we need to know the percentage of tasks dropped reactively (upon missing deadline) and proactively. Our analysis shows that after applying proactive task dropping mechanism, only around 7\% of the task droppings happen reactively. This indicates that proactive task dropping is remarkably effective in avoiding resource wastage and allocates tasks to machines, only if they can complete on time. Proactively dropping tasks with a low chance of success offers a higher chance and certainty of success to the remaining tasks, hence, improving the system robustness. %The analysis also reveals that proactive dropping mechanism makes the reactive dropping mechanism idle.

\subsection{Analysis of the Incurred Cost of using Resources}
While the focus of this paper is to maximize the system robustness in an HC system, there are other metrics of success to consider; one of these is cost. Time consumed for computing tasks that eventually fail to complete on time is a resource wastage that for certain scenarios, such as cloud computing, have associated costs. As such, the aim of this experiment is to analyse the impact of proactive task dropping heuristic on the incurred cost of using such resources. For that purpose, pricing from Amazon cloud~\cite{aws} was mapped to the simulation machines. To create a normalized view of the incurred costs, the price incurred to process the tasks is divided by the percentage of tasks completed on time. We conduct this experiment for various oversubscription levels.

\begin{figure} [h]
  \centering
  \includegraphics[width=0.34\textwidth]{\paperfolder/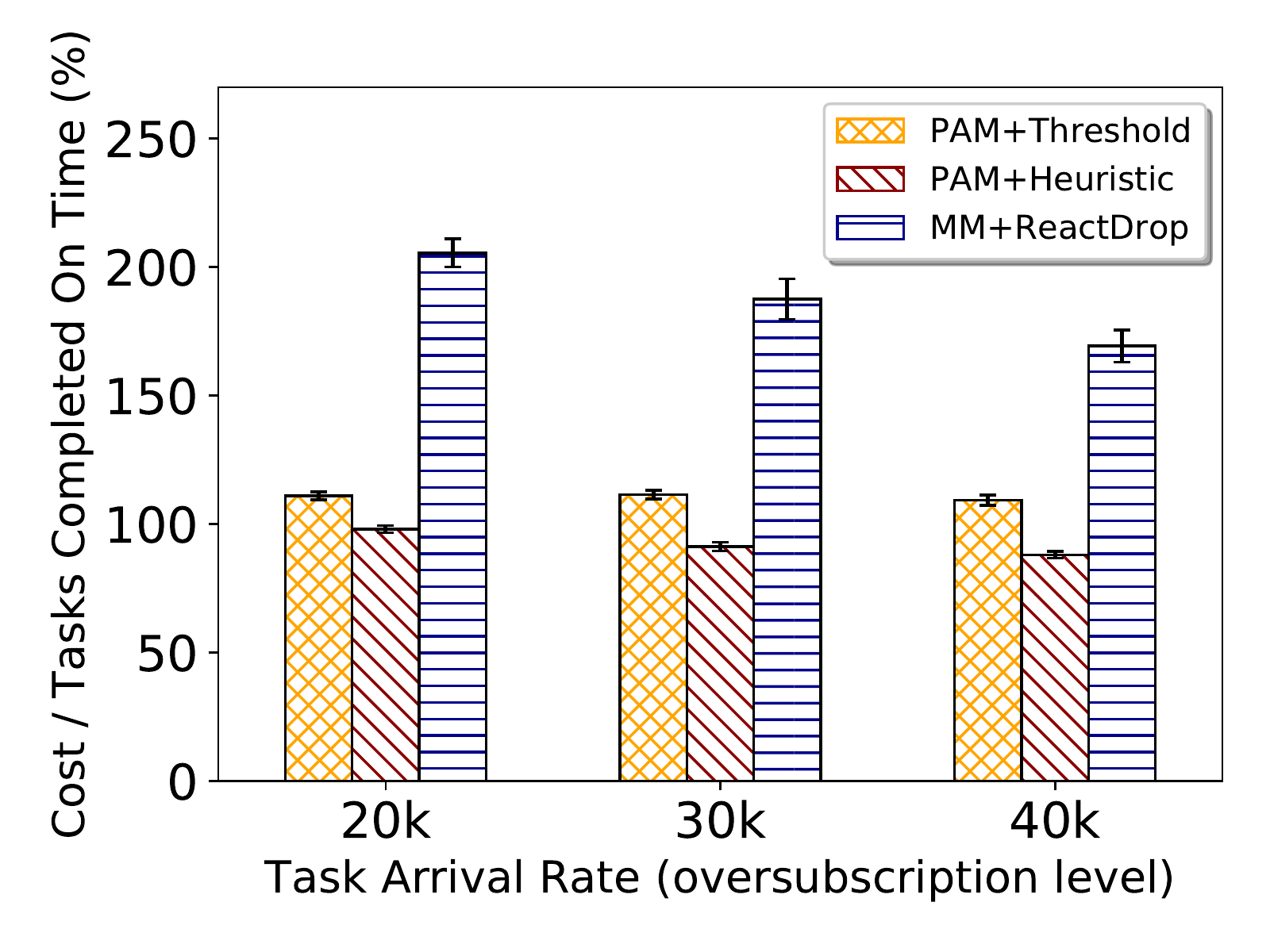}
  \caption{The impact of the proactive task dropping on incurred costs of using resources. Horizontal axis shows the oversubscription level. \label{cost} }
  \vspace{-3px}
\end{figure}

Figure \ref{cost} shows that in an oversubscribed system both PAM+Threshold and PAM+Heuristic incur a significantly ($\simeq$50\%) lower cost per completed task than MM.
In particular, the reason for the improvement in PAM+Heuristic is prioritizing tasks that are most likely to succeed. The significance of this experiment is showing the fact that PAM+Heuristic not only outperforms other dropping-based methods in terms of robustness, but it also performs that with a lower incurred cost, because of not processing tasks needlessly.

\subsection{Validating Robustness for Video Transcoding Workload}

\begin{figure}[h]
  \centering
  \includegraphics[width=0.34\textwidth]{\paperfolder/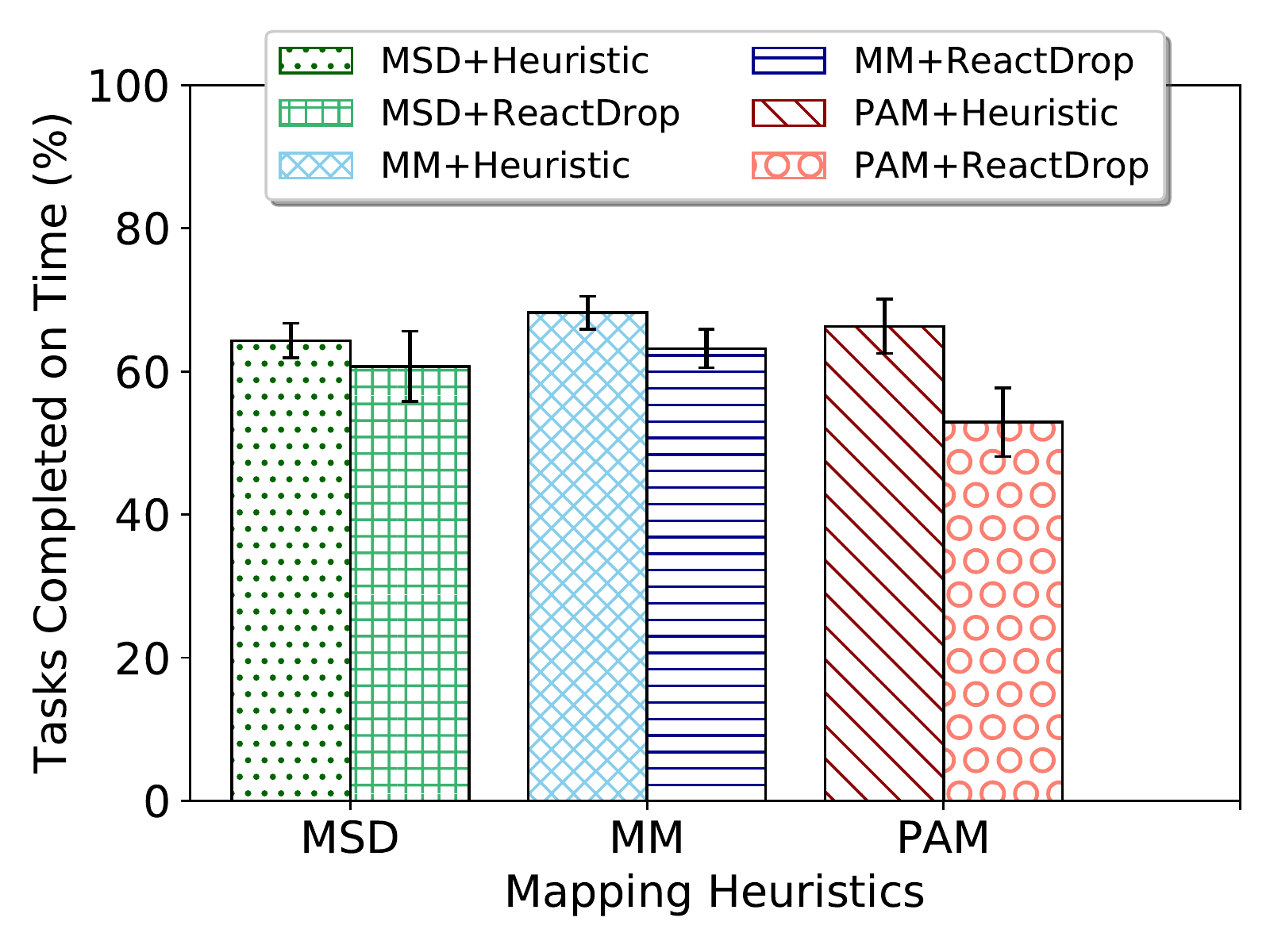}
  \caption{The impact of proactive task dropping applied on the video transcoding workload using different mapping heuristics. Oversubscription level is 20k tasks. \label{fig:VideoWorkload} }
\end{figure}

To validate our earlier observations, we utilize video transcoding workload traces to measure the impact of proactive dropping heuristic on the system robustness. The video workload includes four video transcoding (task) types on four heterogeneous machine types (two machines for each type).  %discrepancy of task-machine affinity is reduced (\ie more consistent heterogeneity) comparing to the workload trials used in other experiments. However, the 
Execution time variation across different task types is high (\ie certain task type takes significantly shorter time to execute than the others across all machine types). These video workload traces also have a lower arrival rate and the system is moderately oversubscribed.

The results, shown in Figure~\ref{fig:VideoWorkload}, confirms our earlier observations that applying proactive task dropping heuristic improves the system robustness, regardless of the mapping heuristic deployed in the system. Further, we observe that when proactive task dropping is plugged into the system, all mapping heuristics exhibit almost the same robustness, which again validates our observations in the earlier experiments. 

\section{Conclusion and Future works}
\label{sec:conclsn}
In this paper, we investigated robustness of HC systems against the compound uncertainty resulted from both uncertain task execution times and uncertain task arrivals. To attain the robustness goal, we proposed an autonomous dropping mechanism that captures the compound uncertainty and proactively drops tasks whose chance of success is low, to increase the chance of success for the remaining tasks, hence, maximizing the overall system robustness. The dropping mechanism uses a mathematical model to determine the optimal task dropping decisions in a dynamic resource allocation system. We then utilized the mathematical model and proposed a sub-optimal task dropping heuristic that provides nearly the same robustness as the optimal one. Experimental results show that the proactive task dropping heuristic not only improves the system robustness in both heterogeneous and homogeneous systems by around 20\%, but also reduces the incurred cost of using resources. In compare to earlier task dropping works, the proposed proactive task dropping mechanism provides the following advantages: (A) It is dynamic and does not require user intervention to configure any predetermined threshold;  (B) Architecturally, it is less complicated and can cooperate with any mapping heuristic in a resource allocation system; (C) It provides a higher system robustness. 

In future, we plan to extend the probabilistic analysis to consider approximately computing tasks, in addition to task dropping. Finally, we plan to extend the probabilistic analysis and cover other types of compound uncertainties, such as those resulted from network latency and resource failure.

%one paper mentioned in multiple sections, each talking in one aspect
\section*{Acknowledgments}
We would like to thank the anonymous reviewers of the paper.
Portions of this research were conducted with high performance computational resources provided by the Louisiana Optical Network Infrastructure (http://www.loni.org).
This research was supported by the Louisiana Board of Regents under grant number LEQSF(2016-19)-RD-A-25. 

\bibliographystyle{IEEEtran}
\balance
\bibliography{paper}
\end{document}